\documentclass[a4paper,prl,twocolumn,showpacs,superscriptaddress,floatfix,nofootinbib]{revtex4-1}
\usepackage{times}
\usepackage{amssymb,amsmath}

\newcommand{\di}{\mathrm{d}}

\newcommand{\vth}{\vartheta}

\usepackage{graphicx}
\usepackage{dcolumn}
\usepackage{bm}
\usepackage{xcolor}
\definecolor{navy}{RGB}{0,0,150}
\usepackage[colorlinks=true,allcolors=navy]{hyperref}
\usepackage{fancyhdr}
\begin{document}

\preprint{APS/123-QED}

\title{Spin-perturbed orbits near black holes}

\author{Vojt{\v e}ch Witzany}
\email{vojtech.witzany@asu.cas.cz}
\affiliation{%
Astronomical  Institute  of  the  Academy  of  Sciences  of  the  Czech  Republic,
Bo\v{c}n{\'i}  II  1401/1a,  CZ-141  00  Prague,  Czech  Republic
}%
\affiliation{%
Center of Applied Space Technology and Microgravity (ZARM), 
Universit{\"a}t Bremen, 
Am Fallturm 2,  
D-28359 Bremen, Germany
}%

\date{\today}

\begin{abstract}
Inspirals of rotating stellar-mass compact objects into massive black holes are influenced by the spin-curvature coupling, which drives the compact body away from geodesic motion due to its rotation. I formulate the Hamilton-Jacobi equation for a spinning test body orbiting a Kerr black hole, solve it to linear order in spin, and use the perturbative solution to compute the fundamental frequencies of motion along the orbit. This result provides one of the necessary ingredients for waveform models for the upcoming space-based gravitational-wave detector LISA. 
\end{abstract}

\maketitle
\newpage



\noindent{\em Introduction} Gravitational-wave inspirals of stellar-mass compact objects into massive black holes are expected to be one of the key sources of gravitational radiation for the upcoming space-based detector LISA \citep{amaro2017}. These so-called extreme mass ratio inspirals (EMRIs) are most accurately treated in a post-geodesic expansion, where the locally almost geodesic motion of the small compact object receives corrections from radiation-reaction and finite-size effects \citep{poisson2011,harte2012,steinhoff2012,barack2018}. The leading order finite-size effect is the so-called spin-curvature coupling, which can be understood as a consequence of the interaction of the gravitomagnetic dipole of the compact object with the field of the massive black hole. A waveform model for EMRIs that will have an accurate phase throughout the entire inspiral requires the inclusion of radiation-reaction forces on the orbit valid up to second order in the mass of the compact object \citep{pound2012,gralla2012,pound2017}, the spin-curvature coupling \citep{huerta2011,huerta2012,Warburton2017,burko2015}, and ``cross-effects'' of the radiation-reaction and finite size \citep{hinderer2008,barack2018}. 

While the computation of the radiation-reaction effects is currently underway \citep{pound2017,vandemeent2018b}, the implementation of finite-size effects in EMRI models has been less intensive. There is a large body of literature studying, mostly numerically, the evolution of test bodies with spin in unperturbed black-hole space-times (see \citep{semerak1999,kyrsem} for a review), and some of those have computed changes to outgoing gravitational waves caused by the spin perturbation to the motion \citep{tanaka1996,suzuki1997,tominaga2001,Harms2016,lukes2017}. Another thread of research focused on the evolution of the small body spin along an orbit under the radiation-reaction force \citep{dolan2014,akcay2017,kavanagh2017,akcay2017b,bini2018}, even though the main application of such computations is not in EMRI models but in checking and refining the results of other approaches to the relativistic two-body problem. Finally, the only works to date that carried out concrete computations of EMRIs of bodies with spin are Refs. \citep{huerta2011,huerta2012,burko2015,will2017,Warburton2017}.

In this letter, I present a semi-analytical scheme of obtaining spin corrections to observables of generic orbits in Kerr space-time (the field of an isolated rotating black hole in Einstein gravity). In particular, these corrections can be readily implemented in EMRI models based on the two-timescale or averaging methods \citep{hinderer2008,vandemeent2018}. The core of the scheme is an analytical perturbative solution of the Hamilton-Jacobi equation for a spinning body based on the Hamiltonian formalism recently obtained in \citep{spinpap}. The solution is separable away from turning points and the separation constants are the R{\"u}diger constants \citep{ruediger1,ruediger2} that arise thanks to the ``hidden symmetry'' of Kerr space-time.

The solution of the Hamilton-Jacobi equation reduces the order of the equations of motion by half, but does not allow for a full separation of the orbital equations of motion. However, it still allows to analytically compute the turning points of the orbits, and to provide closed integral formulas for the fundamental frequencies of the spin-perturbed motion. The results also provide important insights for the interaction of the spin with radiation reaction. 

I use the $G=c=1$ geometrized units and the (-+++) signature of the metric. The metric $g_{\mu\nu}$ used in this paper is always the Kerr metric in Boyer-Lindquist coordinates $t,\varphi,r,\vartheta$ \citep{boyer1967} with black-hole spin parameter $a$ and mass $M$. I will often use $y$ as a placeholder for either $r$ or $\vartheta$. Greek indices label coordinate components and large Latin indices label tetrad components; both run from 0 to 3. A comma before an index signifies an ordinary derivative with respect to the coordinate, a semi-colon a covariant derivative. Additional details will appear in a follow-up paper \citep{spinpertPRD}.


\noindent{\em Hamiltonian and canonical coordinates} The state of the spinning body is approximately characterized by the position of its center of mass $x^\mu$, specific momentum $U_\mu$, and specific spin $s^{\mu\nu}=-s^{\nu\mu}$ \citep{mathisson1937,papapetrou1951,dixon1964}. Higher order mass-multipole moments are neglected here. Under the Tulczyjew-Dixon \citep{tulczyjew1959,dixon1970} frame condition $s^{\mu\nu}U_\mu=0$ the Hamiltonian for the motion of the spinning body reads \citep{spinpap}
\begin{subequations} \label{eq:tdham}
\begin{align}
& H_\mathrm{TD} = \frac{1}{2}  \left( g^{\mu\nu} - \gamma^{\mu\nu}\right) U_\mu U_\nu  \,, \\
& \gamma^{\mu\nu} \equiv \frac{4 s^{\nu\gamma} R^\mu_{\; \gamma \kappa \lambda} s^{\kappa \lambda} }{4 +  R_{\chi \eta \omega \xi} s^{\chi \eta} s^{\omega \xi}}  \,,
\end{align}
\end{subequations}
where on shell $H_{\rm TD} = -1/2$.
The canonical coordinates on the phase space are the pairs $x^\mu,\mathcal{U}_\mu;\phi,\mathcal{A};\psi,\mathcal{B}$, where $\mathcal{U}_\mu = U_\mu + e_{C\nu;\mu}e^\nu_D s^{CD}/2$, $e_{C}^\kappa$ is some orthonormal tetrad, and $s^{CD}$ tetrad components of spin. The coordinates $\phi,\mathcal{A};\psi,\mathcal{B}$ then parametrize the spin tensor, the transformation $s^{CD} = s^{CD}(\phi,\psi,\mathcal{A},\mathcal{B})$ is given in \citep{spinpap}. The important properties are that $s^{12}=-s^{21} = \mathcal{A}+\mathcal{B}+s$, where $s\equiv\sqrt{s^{CD}s_{CD}/2}$ is a constant of motion. Furthermore, $s^{0D} = 0$ implies $\mathcal{B} = 0$ and that $\psi$ becomes a non-consequential coordinate.


\noindent{\em Adapted tetrad} The choice of the tetrad ``orients'' the base of the canonical coordinates and will be essential in solving the Hamilton-Jacobi equation. I choose the zeroth tetrad leg to be a geodesic congruence in Kerr space-time $e_{0\mu} = u^{\rm c}_\mu$. The congruence is specified by signs of $u^{\rm c}_r,u^{\rm c}_\vartheta$ and constants of motion $K_{\rm c},E_{\rm c},L_{\rm c}$, where $K_{\rm c}$ is the Carter constant \citep{carter1968}, $E_{\rm c}$ the specific energy, and $L_{\rm c}$ specific azimuthal angular momentum. The other three legs are generated with the help of the so-called Killing-Yano tensor $Y_{\mu\nu}=-Y_{\nu\mu},Y_{\mu\nu;\kappa} = - Y_{\mu\kappa;\nu}$ \citep{floyd1973,penrose1973}. I define $e_{3\mu} = Y_{\mu\nu}u^{{\rm c}\nu}/\sqrt{K_{\rm c}}$, and $e_{1\mu},e_{2\mu}$ are then obtained by multiplying $e_{3\mu}$ by one and two additional powers of $Y_{\mu\nu}$ respectively, and orthogonalizing with respect to $e_{0\mu},e_{3\mu}$ (the orthogonalization turns out to be unique and the tetrad equivalent to the one constructed by different means by \citet{marck}). Note that the tetrad is defined only within the turning points of the geodesic congruence $u^{\rm c}_\mu$ and the spin connection will be singular there. 

When the dust settles, the only component of the spin connection that has a non-zero projection in the zeroth leg reads \citep{marck}
\begin{align}
\begin{split}
    & e_{2;\mu}^\kappa e_{1\kappa} e^{\mu}_0 = 
    \\
    & \frac{\sqrt{K_\mathrm{c}}}{r^2 + a^2 \cos^2  \! \vartheta} \Big( \frac{E_\mathrm{c} (r^2 + a^2) - a L_\mathrm{c}}{ r^2 + K_\mathrm{c}} + a \frac{L_\mathrm{c} - a E_\mathrm{c} \sin^2 \! \vartheta}{K_\mathrm{c} -  a^2 \cos^2 \! \vartheta} \Big)\,,\label{eq:marckform}
\end{split}
\end{align}
where $a$ is the spin parameter of the black hole. I will choose the zeroth leg to be close to the vector $U_\mu$, so, up to higher order terms, $s^{0D} = 0$, and thus also $\mathcal{B} = 0$. The spin tensor then simplifies as $s^{12} = \mathcal{A} +s,\,s^{23} = \sqrt{\mathcal{A}(\mathcal{A} + 2s)} \sin \phi,\,s^{31} = \sqrt{\mathcal{A}(\mathcal{A} + 2s)} \cos \phi$.


\noindent{\em Hamilton-Jacobi equation} The Hamilton-Jacobi equation for the function $W(t,\varphi,r,\vartheta,\phi,\psi)$ is constructed by replacing any of the canonical momenta in \eqref{eq:tdham} with gradients with respect to their canonically conjugate coordinates. Away from turning points of the background congruence, or in the ``swing region'' the Hamilton-Jacobi equation up to $\mathcal{O}(s)$ terms reads
\begin{align}
g^{\mu\nu} W^{(1 \rm sw)}_{,\mu} W^{(1 \rm sw)}_{,\nu}  - e^\kappa_{C;\nu}e_{D\kappa } s^{CD} W^{(0),\nu} + \mathcal{O}(s^2) =-1 \,, \label{eq:pertspjac}
\end{align}
where $W^{(0)}$ is the zeroth-order (geodesic) solution for the action given by \citet{carter1968} and $W^{(1\rm sw)}$ the first order action in the swing region. I now assume that $W^{(0)}_{,\mu}$ is chosen $\mathcal{O}(s)$ close to $u_\mu^{\rm c}$, the perturbing term in \eqref{eq:pertspjac} then simplifies due to \eqref{eq:marckform} and $\phi$ becomes a cyclical coordinate. Finally, I use a separable Ansatz $W^{(1 \rm sw)} = -E_{\rm so}t + L_{\rm so} \varphi + (s_{\parallel} -s)\phi + w_{r}(r) + w_\vartheta(\vartheta)$ to obtain
\begin{subequations}
\begin{align}
\begin{split}
(w'_\vartheta)^2 =& K_\mathrm{so} - \left( \frac{L_\mathrm{so}}{\sin \vartheta} - a E_\mathrm{so} \sin \vartheta \right)^2 - a^2 \cos^2 \! \vartheta \\  & -2a s_{\parallel}\sqrt{K_\mathrm{c}} \frac{L_\mathrm{c} - a E_\mathrm{c} \sin^2 \! \vartheta}{K_\mathrm{c} -  a^2 \cos^2  \! \vartheta}  \,,\end{split} \label{eq:wth}
\\
\begin{split}
\Delta (w'_r)^2 =& -K_\mathrm{so} + \frac{1}{\Delta} \left(E_\mathrm{so} (r^2 + a^2) - a  L_\mathrm{so} \right)^2 -  r^2 \\  & -2  s_{\parallel} \sqrt{K_\mathrm{c}} \frac{ E_\mathrm{c} (r^2 + a^2) - a L_\mathrm{c}}{K_\mathrm{c}+ r^2} \,, \end{split}\label{eq:wr}
\end{align}
\end{subequations}
where $s_{\parallel},K_{\rm so}, E_{\rm so}, L_{\rm so}$ are separation constants of the solution and integrals of motion of the spin-perturbed orbit. $K_{\rm so}$ and $s_{\parallel}$ are conserved thanks to the existence of the Killing-Yano tensor and are equivalent to the constants of motion for spinning particles found by \citet{ruediger1,ruediger2}. $K_{\rm so}$ is interpreted as specific spin-orbital angular momentum squared and $s_{\parallel}$ as the component of spin aligned with orbital angular momentum.

However, the Hamilton-Jacobi equation receives other corrections whenever approaching a turning point of the background congruence $y_{\rm c t}$ ($y=r,\vartheta$) because of singular coordinate dependencies of four-velocities at such points. When $y-y_{\rm ct} = \mathcal{O}(s)$, we are in the ``$y$-turning region'' and all the relevant terms are
\begin{align}
\begin{split}
&g^{\mu\nu} W^{(1)}_{,\mu} W^{(1)}_{,\nu}  - e^\kappa_{C;\nu}e_{D \kappa} s^{CD}e_0^\nu+1
\\
& \left[- e^\kappa_{C;y}e_{D \kappa} s^{CD}(W^{(1)}_{,y} - e_{0 y}) + \frac{1}{4} (e_{C;y}^\kappa e_{D \kappa} s^{CD})^2\right] g^{yy} 
\\&=0 + \mathcal{O}(s^2) \,. 
\end{split}\label{eq:turnjac}
\end{align}
By assuming that $W^{\rm (1 sw)}$ receives corrections that become large only in their respective $y$-turning regions, it is possible to derive the following action that fulfills the Hamilton-Jacobi equation with at most $\mathcal{O}(s^2)$ residual terms in swing regions and $\mathcal{O}(s^{3/2})$ residual terms in turning regions
\begin{align}
\begin{split}
    W^{(1)}(t,\varphi,r,\vartheta,\phi) =& (s_\parallel - s) \phi -E_{\rm so} t + L_{\rm so} \varphi \\ 
    + \sum_{y=r,\vartheta} \int \Big(& \pm \sqrt{w_y'^2 -   e_{0y}{e}_{C;y}^\kappa  {e}_{D \kappa} \tilde{s}^{CD}} \\ 
    & +  \frac{1}{2}  {e}_{C;y}^\kappa  {e}_{D \kappa} \tilde{s}^{CD}  \Big)\di y \,, \label{eq:wfull}
\end{split}
\end{align}
where $\tilde{s}^{CD} = -\tilde{s}^{DC}, \,\tilde{s}^{0D}=0,\,\tilde{s}^{12}= s_\parallel,\,\tilde{s}^{23} = \sqrt{s^2 - s_\parallel^2} \sin \phi,\, \tilde{s}^{31} = \sqrt{s^2 - s_\parallel^2} \cos \phi$. 

First, note that $\psi$ does not appear in the action because the condition $s^{\mu\nu}U_\mu=0$ makes it redundant at given order. Second, note that this construction works only if we can choose $K_{\rm c},E_{\rm c},L_{\rm c}$ $\mathcal{O}(s)$-close to $K_{\rm so},E_{\rm so},L_{\rm so}$ while making the turning points of the tetrad congruence wider than the turning points of the spin-perturbed orbit. This will not be possible when very close to transitions between bound and unbound (plunging) motion.


\begin{figure*}
    \centering
    \includegraphics[width=0.4\textwidth]{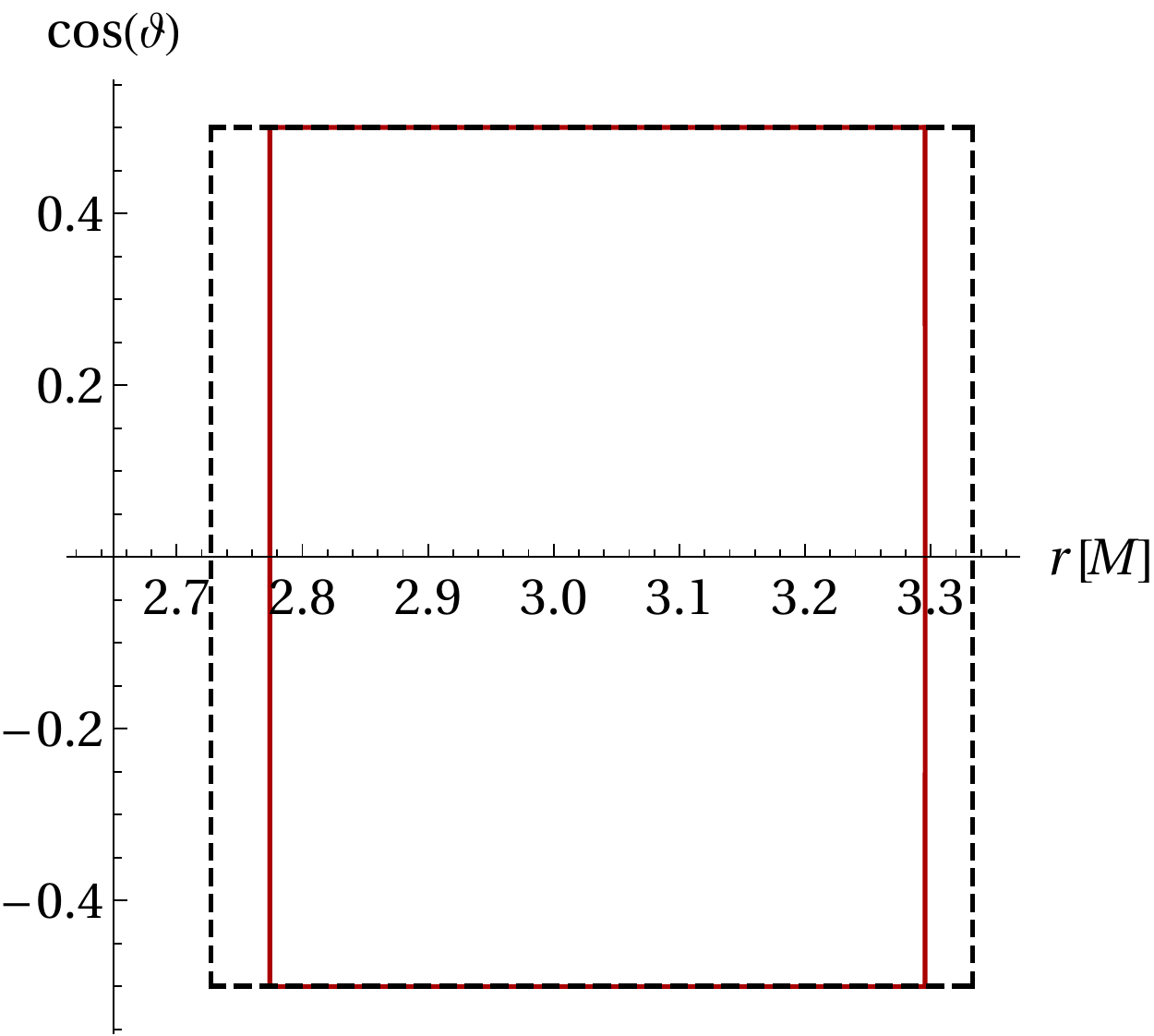} \includegraphics[width=0.4\textwidth]{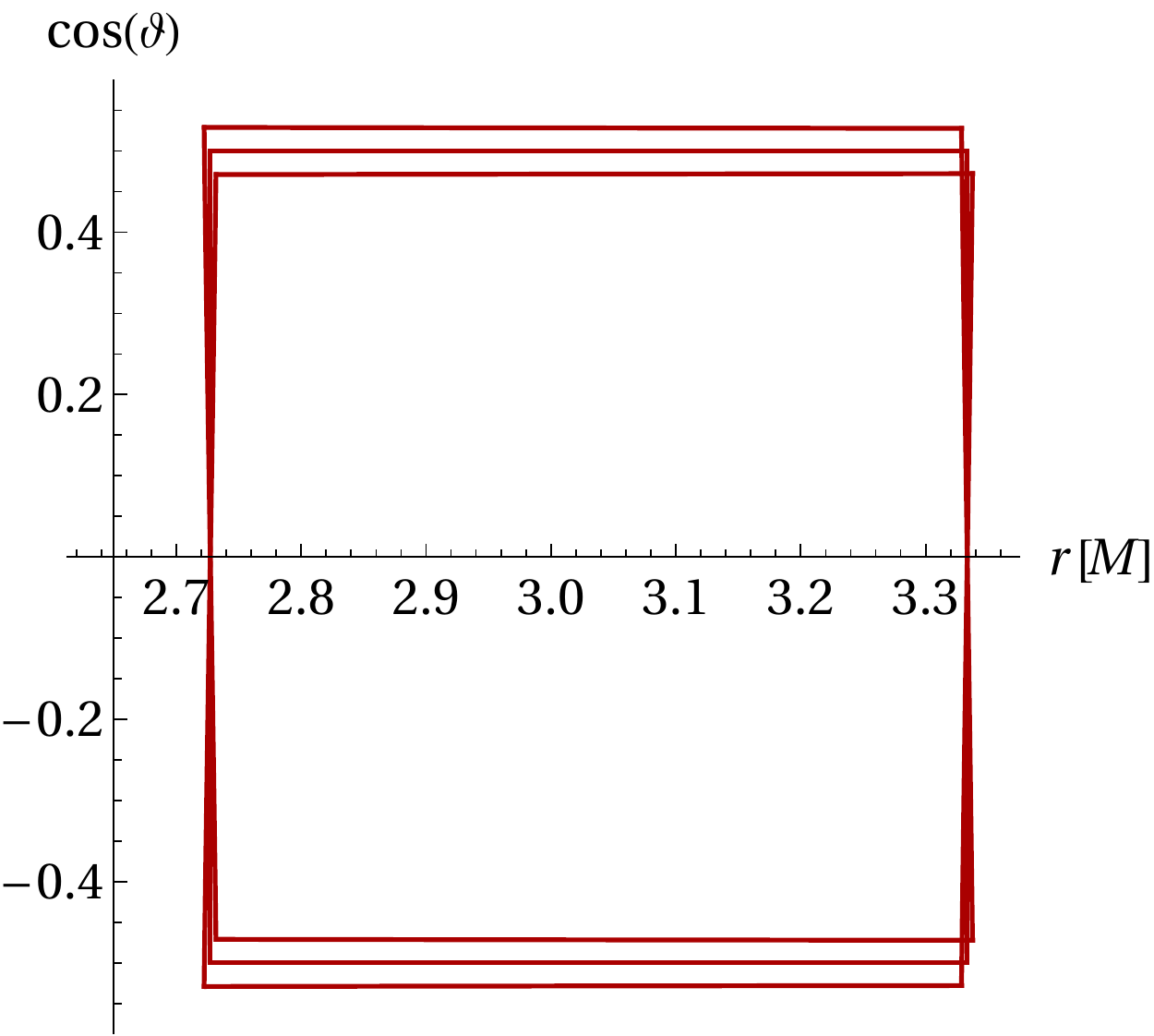}
    \caption{The turning points of bodies near a Kerr black hole with $a=0.9M$ and with various choices of the body spin. The fiducial geodesic is always with $K = 2.9 M^2, E = 0.87 , L = 2.0 M$ (or semi-latus rectum $p=3M$, eccentricity $e = 0.1$, and inclination $\cos \vartheta_{\rm min} = 0.5 $, see \citep{schmidt2002,drasco2004}). On the left I show the turning points of the fiducial geodesic (dotted black) and the turning points of a spin-perturbed orbit with completely aligned spin, $s_\parallel = s = 10^{-3} M$ (full dark red). On the right I show the turning points of an orbit with a completely oscillating spin $s_\parallel=0, s= 5\cdot 10^{-2}   M$ for the values of spin angle $\phi=0,\pi/2,3\pi/2$. }
    \label{fig:turnboxes}
\end{figure*}

\noindent{\em Orbital shape} The equations of motion corresponding to \eqref{eq:wfull} read
\begin{subequations}
\label{eq:vel}
\begin{align}
 \frac{\di r}{\di \lambda} = &\pm\Delta\sqrt{{w'_r}^2 - e_{0r}  {e}^\kappa_{C;r} {e}_{\kappa B}  \tilde{s}^{CD} } \,, \\
 \frac{\di \vartheta}{\di \lambda} = &\pm  \sqrt{{w'_\vartheta}^2 -  e_{0 \vartheta}  {e}^\kappa_{C;\vartheta} {e}_{\kappa B}  \tilde{s}^{CD} }\,, \\
\begin{split}
\frac{\di \phi}{\di \lambda} =  
&-\sqrt{K_\mathrm{c}} \Big( \frac{E _\mathrm{c} (r^2 + a^2) - a L _\mathrm{c}}{K_\mathrm{c}+r^2}
\\
&\quad\quad\quad+ a \frac{L _\mathrm{c} - a E _\mathrm{c} \sin^2 \! \vartheta}{K _\mathrm{c} - a^2 \cos^2 \! \vartheta} \Big)\,,\label{eq:phipr}
\end{split}\\
\Delta \equiv & \,r^2 - 2Mr +a^2\,,
\end{align}
\end{subequations}
where $\mathrm{d} \lambda = \mathrm{d}\tau/(r^2 + a^2 \cos^2 \! \vartheta)$ is the Mino time \citep{mino2003}, and terms were discarded so that the $r,\vartheta$ orbital shape is known up to $\mathcal{O}(s)$. The equations \eqref{eq:vel} are not separable because of the connection terms and the appearance of $\phi$ in $\tilde{s}^{CD}$. Furthermore, it can be shown that symmetry of the equations of motion with respect to reflections about the equatorial plane appears only in a generalized sense.

However, it is still possible to find turning points analytically by assuming they are $\mathcal{O}(s)$ close to the turning points $y_{\rm gt}$ of a fiducial geodesic, $y_{\rm t} = y_{\rm gt} + \delta y_{\rm t}$. When one chooses the fiducial geodesic parameters as $K=K_\mathrm{so} - 2 a s_\parallel \mathrm{sgn}(L_\mathrm{so} - a E_\mathrm{so})$, $L=L_\mathrm{so}$, and $E=E_\mathrm{so}$, the turning point shifts read
\begin{subequations} \label{eq:turnshifts}
\begin{align}
     \delta r_\mathrm{t} =& \frac{2 s_\parallel \mathcal{G} + \Delta (K + r^2 )  e_{0r} e_{C \kappa;r}e_{D}^\kappa \tilde{s}^{CD}}{\Upsilon_{(r)} (K +  r^2)}\Big|_{r=r_\mathrm{gt}} \!\!, \label{eq:deltart}
    \\
     \delta \vartheta_\mathrm{t} =& \frac{2as_\parallel \mathcal{H} + (K  -   a^2 \cos^2\! \vartheta) e_{0\vartheta} e_{C \kappa;\vartheta}e_{D}^\kappa \tilde{s}^{CD}}{\Upsilon_{(\vartheta)}(K  -   a^2 \cos^2\! \vartheta)}\Big|_{\vartheta=\vartheta_\mathrm{gt}} \!\!, \label{eq:deltatht}
    \\
    \begin{split}
    \mathcal{G} \equiv& \sqrt{K}(E(r^2 + a^2) -a L ) \\
    &+ a (K +   r^2) \mathrm{sgn}(L-aE)\,,
    \end{split}\\
    \mathcal{I}_{(r)} \equiv&  \frac{\di}{\di r}\left(\Delta^{-1} \left[E (r^2 + a^2) - a L \right]^2 -   r^2\right) \,, \\
    \begin{split}
     \mathcal{H} \equiv& \sqrt{K } (L  - a E  \sin^2\!\vartheta) 
    \\ & - (K  -   a^2 \cos^2\! \vartheta) \mathrm{sgn}(L-aE) \,,
    \end{split}\\
     \mathcal{I}_{(\vartheta)} \equiv & -\frac{\di}{\di \vartheta} \left[ \left(L - a E \sin^2\! \vartheta \right)^2 \sin^{-2} \! \vartheta +   a^2 \cos^2 \! \vartheta \right] \!.
\end{align}
\end{subequations}
Sample ``turning boxes'' are plotted in Fig. \ref{fig:turnboxes}. The choice $K=K_\mathrm{so} - 2 a s_\parallel \mathrm{sgn}(L_\mathrm{so} - a E_\mathrm{so})$ ensures that the fiducial geodesic stays $\mathcal{O}(s)$ close to the spin-perturbed motion even as it gets restricted to the equatorial plane. However, the fiducial geodesics chosen here become $\mathcal{O}(\sqrt{s})$ far from the spin-perturbed orbits for near-circular motion and for such cases the above shifts diverge. 



\begin{figure}
    \centering
    \includegraphics[width=0.48\textwidth]{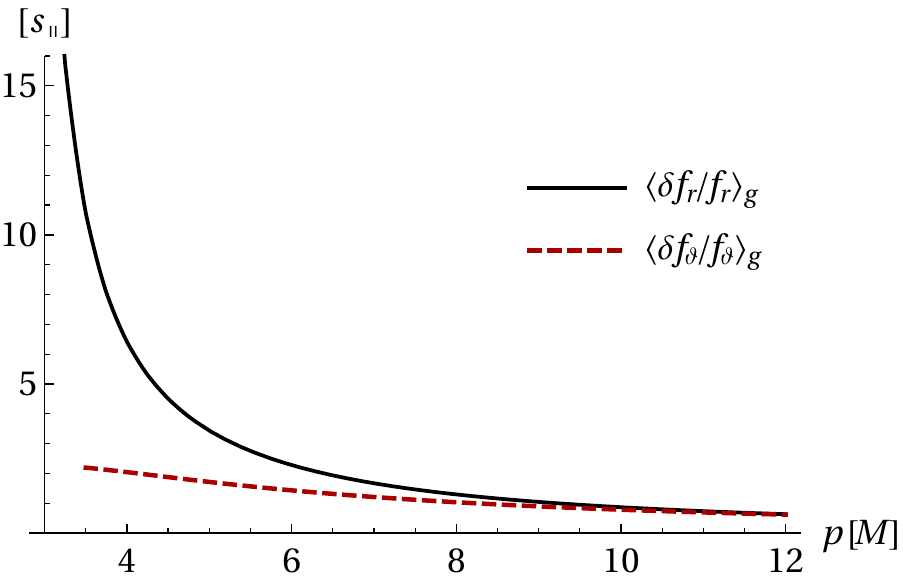}
    \caption{The relative corrections to fundamental frequencies given in units of the aligned component of spin $s_\parallel$ as a function of semi-latus rectum $p$ for fiducial geodesics with eccentricity $e=0.1$, inclination $z_{- \rm g} = 0.1$, and $a=0.9 M$ (see \citep{drasco2004} for definitions of the orbital parameters $p,e,z_{-\rm g}$). The corrections to the radial frequency becomes large at small $p$ because the respective orbits are closer to the black holes and thus less radially stable.}
    \label{fig:freqs}
\end{figure}


\noindent{\em Frequencies of motion} Consider a transformation to angle-type coordinates $\chi_r,\chi_\vartheta$ such that
\begin{subequations} \label{eq:angparam}
\begin{align}
    & r(\chi_r,\chi_\vartheta,\phi) = r_0 + \delta r \,, \label{eq:rparam}\\
    & \cos \left[ \vartheta(\chi_r,\chi_\vartheta,\phi) \right] = \cos \vartheta_0 - \delta \vartheta \sqrt{1 - z_{- \rm g}} \,,\label{eq:thparam}\\
    & r_0 = \frac{r_{1\rm g}  + r_{2\rm g}}{2} + \frac{r_{1\rm g}  - r_{2\rm g}}{2} \sin\chi_r \,,\\
    & \delta r = \frac{\delta r_{1}  + \delta r_{2}}{2} + \frac{\delta r_{1}  - \delta r_{2}}{2} \sin\chi_r \,,\\
    & \cos\vartheta_0  = \sqrt{z_{-\rm g}} \sin\chi_\vartheta \,,\\
    & \delta \vartheta = \frac{\delta \vartheta_{1}  + \delta \vartheta_{2}}{2} + \frac{\delta \vartheta_{1}  - \delta \vartheta_{2}}{2} \sin\chi_\vartheta\,,\\
    & \delta y_i(\chi_x,\phi) \equiv \delta y_{\rm t} (y_{\rm g i},x_0(\chi_x),\phi)\,,
\end{align}
\end{subequations}
where $x,y$ is either $r,\vartheta$ or $\vartheta,r$, and $z_{-\rm g}$ is the minimum value of $\cos^2\! \vartheta$ of the fiducial geodesic \citep{drasco2004}. Then the equations of motion acquire the general form
\begin{align}
    & \frac{\di \chi_y}{\di \lambda} = f_{y}(\chi_y) + \delta f_{y}(\chi_r,\chi_\vartheta,\phi)\,,\\
    & \frac{\di \phi}{\di \lambda} = h_r(\chi_r) + h_\vartheta(\chi_\vth)\,, \label{eq:phip}
\end{align}
where $\delta f_y$ are $\mathcal{O}(s)$ and the rest of the terms are $\mathcal{O}(1)$. Any state of the trajectory is now given by some $\chi_r,\chi_\vartheta,\phi \in [0,2\pi)^3$ and it is easy to apply usual perturbation theory to obtain the fundamental frequencies of the system as (e.g. \citep{arnold2007})
\begin{subequations}
\begin{align}
    &\Upsilon_y = \Upsilon_{y\rm g}\left(1 + \left\langle \frac{\delta f_y}{f_y} \right\rangle_{\rm \!\!g}\right)\,,
    \\
    & \langle j(\chi_r,\chi_\vartheta,\phi) \rangle_{\rm g}\equiv \frac{\Upsilon_{r \rm g} \Upsilon_{\vartheta \rm g}}{(2\pi)^3} \int_{[0,2\pi)^3}\!\! \frac{j\, \mathrm{d}\chi_r \mathrm{d}\chi_\vartheta \mathrm{d}\phi}{f_r f_\vartheta}\,,
\end{align}
\end{subequations}
where $\Upsilon_{y \rm g}$ are the fundamental Mino frequencies of the fiducial geodesic as given by \citet{fujita2009}. The fundamental Mino frequency of $\phi$ is obtained by simply taking the geodesic average of \eqref{eq:phip} and it will be expressible in terms of special functions thanks to its separability. The computation of the fundamental frequencies is done under the assumption of non-resonant motion, that is, under the assumption that none of the  fundamental frequencies are in an integer ratio. A sample of concrete values of $\langle \delta f_y/f_y\rangle_{\rm g}$ is given in fig. \ref{fig:freqs}. 

To obtain observables such as coordinate-time frequencies one also needs to compute the averages of $\mathrm{d}t/\mathrm{d}\lambda$ and $\mathrm{d}\varphi/\mathrm{d}\lambda$ over the spin-perturbed orbit, which read
\begin{align}
\begin{split}
    &\Xi \equiv \left\langle \frac{\di t}{\di \lambda}\right\rangle_{\rm \!\! sp} =
    \left\langle\frac{(r^2 + a^2) \mathcal{J}(r)}{\Delta} \right\rangle_{\rm \!\! sp} 
   \\& \quad\quad\quad\quad\quad \quad\; - a E_{\rm so} \left\langle \sin^2 \vartheta\right\rangle_{\rm  sp} +  a L_{\rm so}
   \\& \quad\quad\quad\quad\quad \quad\;- \langle \Gamma_{CD}^{\quad\;t}\,\tilde{s}^{CD} \rangle_{\rm g}
   \,,
    \end{split}\\
    \begin{split}
    &\bar{\Upsilon}_\varphi \equiv \left\langle \frac{\di \varphi}{\di \lambda}\right\rangle_{\rm \!\! sp} =  \left\langle\frac{a\mathcal{J}(r)}{\Delta} \right\rangle_{\rm \!\! sp} + \left\langle\frac{L_{\rm so}}{\sin^2 \vartheta}\right\rangle_{\rm \!\! sp} -  a E_{\rm so} 
    \\& \quad\quad\quad\quad\quad \quad\quad- \langle \Gamma_{CD}^{\quad\;\varphi}\,\tilde{s}^{CD} \rangle_{\rm g} \,,
\end{split} 
\\  &\mathcal{J}(r)= E_\mathrm{so}(r^2 + a^2) - a L_\mathrm{so}\,,
\\  &\Gamma_{CD}^{\quad\;\mu} = \Gamma_{\lambda\kappa\nu}e^\lambda_{C}e^{\kappa}_{D}g^{\mu\nu}\,,
\end{align}
where $\langle \rangle_{\rm sp}$ means averaging over the spin-perturbed orbit and $\Gamma_{\lambda\kappa\nu}$ is the Christoffel symbol. Additionally, it is possible to substitute only $r_0,\vartheta_0$ from \eqref{eq:angparam} in place of $r,\vartheta$ for the Christoffel terms. Fortunately, for functions of either only $r$ or only $\vartheta$ the spin-perturbed average reduces to
\begin{align}
\begin{split}
    \langle n(y) \rangle_{\rm sp}  =& \left(1+\left\langle \frac{\delta f_y}{f_y} \right\rangle_{\rm \!\!g} \right)\langle n(y_0) \rangle_{\rm g} + \langle n'(y_0) \delta y \rangle_{\rm g}
    \\& - \left\langle n(y_0) \frac{\delta f_y}{f_y} \right\rangle_{\!\! \rm g}\,,
\end{split} \label{eq:spavg}
\end{align}
where the functions $y_0, \delta y$ are given in \eqref{eq:angparam}. From this it is easy to obtain average coordinate-time frequencies such as the $r,\vartheta$ frequencies $\bar{\Omega}_y = \Upsilon_y/\Xi$, or the azimuthal frequency $\bar{\Omega}_\varphi =  \bar{\Upsilon}_\varphi/\Xi$. Note that at this level no knowledge of the full spin-perturbed trajectory is required and all the observables are obtained in terms of closed-form geodesic averages. 


\noindent{\em Discussion} It is not clear what are the consequences of the solution of the Hamilton-Jacobi equation for the occurence of resonances in the spin-perturbed system, that is, the possible non-analytical response of orbits with integer-ratio frequencies to the perturbation \citep{arnold2007}. The Hamilton-Jacobi equations as well as its solution do not have any special behaviour near orbits with resonant frequencies and there is no way a special response to the $\mathcal{O}(s)$ perturbation might arise at that level. On the other hand, a non-analytical response to the $\mathcal{O}(s)$ terms might still occur at the level of the motion governed by \eqref{eq:vel}. It seems plausible that the existence of a full set of approximately conserved integrals of motion will suppress the resonant response to the spin perturbation \citep{ruangsri2016}, but I leave a rigorous discussion of this matter for future work.

It was shown in \citep{grant2015,witzany2017} that systems of colliding particles do not conserve their sum of Carter constants. The pole-dipole approximation of a finite body leads to the same equations of motion for any system, be it colliding or non-colliding, so the conservation of $K_{\rm so}$ is still consistent with this no-go theorem. However, a pole-dipole-quadrupole approximation necessarily contains reference to the internal, usually collisional dynamics of the body through various deformability parameters \citep{steinhoff2012,vines2016}. Thus, I believe there will be no conserved ``total Carter constant'' for general compact bodies in the pole-dipole-quadrupole approximation (or beyond linear-in-spin order).

The construction given in this letter is almost trivial to generalize to Carter's general class of Kerr-NUT-(A)dS space-times in four dimensions \citep{carter1968b}. Likewise, it should be possible to generalize to dimension 5 by using the tetrad found by \citet{connell2008}. There are indications that Kerr-NUT-(A)dS space-times of general dimension have a sufficient number of integrals to separate the Hamilton-Jacobi equation in the manner given above \citep{kubiznak2012}.


\noindent{\em Conclusion} The herein presented results provide all the needed observables for the inclusion of conservative finite-size effects in EMRI waveform models. Namely, a typical two-timescale EMRI model will require the spin corrections to the quantities $\Upsilon_r, \Upsilon_\vartheta, \bar{\Upsilon}_\varphi, \Xi$, which are all provided here through closed-form integrals.

As for the dissipative finite-size effects, it is now also clear that one needs to compute only the average radiative dissipation of $s_\parallel$ to characterize the decay of the internal angular momentum of the light compact object (the change of $s$ is deducible from the fact that there will be a conserved spin magnitude with respect to the effective metric $g_{\mu\nu} + h_{\mu\nu}^{\rm R}$ \citep{detweiler2003,harte2012}). However, it is also necessary to compute the change in the dissipative rates of the orbital constants of motion due to the spin pertubation. This will require reconstructing the full spin perturbation to the orbital shape, a task which can also be simplified by the herein presented results.

I would like to thank Jan Steinhoff, Justin Vines, Maarten van de Meent, and Georgios Lukes-Gerakopoulos for feedback and numerous discussions of the subject. Preliminary versions of the results in this letter were published in my Ph.D. thesis, which was kindly supervised by Claus L{\"a}mmerzahl and supported from a Ph.D. grant of the German Research Foundation (DFG) within Research Training Group 1620 ``Models of Gravity''. I am also grateful for the support from Grant No. GACR-17-06962Y of the Czech Science Foundation. 

\bibliography{literatura}

\end{document}